\documentclass[12pt]{mn2e}
\usepackage{graphicx}
\usepackage{epsf}
\usepackage{lscape}
\usepackage{longtable}
\usepackage{rotating}
\usepackage{color}
\usepackage{url}

\newcommand{\gtrsim}{\ga}

\newcommand{\aap}{A\&A}

\newcommand{\apj}{ApJ}
\newcommand{\apjl}{ApJ}
\newcommand{\apjs}{ApJS}
\newcommand{\mnras}{MNRAS}

\newcommand{\nat}{Nat}

\definecolor{grey}{rgb}{0.7,0.7,0.7}

\begin{document}
\topmargin -0.5in 

\title[Spectral Properties of $z\sim10$ Galaxies]{Quantifying the UV
  continuum slopes of galaxies to
  $z\sim10$ using deep {\em Hubble}+{\em Spitzer}/IRAC observations}

\author[Stephen M. Wilkins et al.]  
{
Stephen M. Wilkins$^{1}$\thanks{E-mail: s.wilkins@sussex.ac.uk}, Rychard J. Bouwens$^{2}$, Pascal A. Oesch$^{3}$, Ivo Labb{\'e}$^{2}$,\newauthor Mark Sargent$^{1}$, Joseph Caruana$^{4}$, Julie Wardlow$^{5}$, Scott Clay$^{1}$ \\ 
$^1$\,Astronomy Centre, Department of Physics and Astronomy, University of Sussex, Brighton, BN1 9QH, UK \\
$^2$\,Leiden Observatory, Leiden University, P.O. Box 9513, 2300 RA Leiden, The Netherlands\\
$^3$\,Yale Center for Astronomy and Astrophysics, Yale University, P.O. Box 208120, New Haven, CT 06520, USA\\
$^4$\,Leibniz-Institut fur Astrophysik, An der Sternwarte 16, D-14482 Potsdam, Germany\\
$^5$\,Dark Cosmology Centre, Niels Bohr Institute, University of Copenhagen, Denmark\\
}

\maketitle

\begin{abstract}

Measurements of the $UV$-continuum slopes $\beta$ provide valuable
information on the physical properties of galaxies forming in the
early universe, probing the dust reddening, age, metal content, and
even the escape fraction.  While constraints on these slopes generally
become more challenging at higher redshifts as the UV continuum shifts
out of the {\em Hubble Space Telescope} bands (particularly at $z>7$),
such a characterisation actually becomes abruptly easier for galaxies
in the redshift window $z=9.5-10.5$ due to the Spitzer/IRAC
3.6$\mu$m-band probing the rest-UV continuum and the long wavelength
baseline between this Spitzer band and the {\em Hubble} H$_{f160w}$
band.  Higher S/N constraints on $\beta$ are possible at $z\sim10$
than at $z=8$.  Here we take advantage of this opportunity and five
recently discovered bright $z=9.5-10.5$ galaxies to present the first
measurements of the mean $\beta$ for a multi-object sample of galaxy
candidates at $z\sim 10$.  We find the measured $\beta_{\rm obs}$'s of
these candidates are $-2.1\pm0.3\pm0.2$ (random and systematic), only
slightly bluer than the measured $\beta$'s ($\beta_{\rm obs}\approx
-1.7$) at $3.5<z<7.5$ for galaxies of similar luminosities.  Small
increases in the stellar ages, metallicities, and dust content of the
galaxy population from $z\sim10$ to $z\sim7$ could easily explain the
apparent evolution in $\beta$.

\end{abstract} 


\section{Introduction}

Thanks to extremely sensitive near-infrared (NIR) imaging obtained
using Wide Field Camera 3 (WFC3) on the {\em Hubble Space Telescope}
it is now possible to routinely identify galaxies \ at very high
redshift ($z>6$: e.g., Oesch et al.\ 2010; Bunker et al.\ 2010;
Wilkins et al.\ 2010; Wilkins et al.\ 2011a; Bouwens et al.\ 2011b;
Finkelstein et al.\ 2010, 2012; Oesch et al.\ 2012; McLure et
al.\ 2013; Schmidt et al.\ 2014) with the first samples now being
identified at $z\sim 10$ and beyond, less than 500 Myr after the Big
Bang (Bouwens et al.\ 2011a; Zheng et al.\ 2012; Ellis et al.\ 2013;
Oesch et al.\ 2013, 2014, 2015b; Zitrin et al.\ 2014; Ishigaki et
al.\ 2015; Bouwens et al.\ 2015b).

One area of significant interest in the study of distant galaxies
regards their spectral characteristics.  At very early times, we might
expect galaxies to potentially have very different SEDs than at latter
times.  In particular, one would expect galaxies to be bluer in terms
of their rest-frame UV colours and UV-optical breaks, both due to a
younger stellar population (Wilkins et al.\ 2013a) and a lower dust
content (e.g., Wilkins et al.\ 2011b; Bouwens et al.\ 2009, 2012;
Finkelstein et al.\ 2012).  In fact, there has been some evidence for
bluer UV colours at high redshift (Lehnert \& Bremer 2003; Papovich et
al.\ 2004; Stanway et al.\ 2005; Bouwens et al.\ 2006, 2009, 2012,
2014a; Wilkins et al.\ 2011b; Finkelstein et al.\ 2012; Kurczynski et
al.\ 2014) though the strength of the evolution at the highest
redshifts has been subject to some debate (Dunlop et al.\ 2012;
Robertson et al.\ 2013).  Evolution in the size of the Balmer break is
less easily inferred (Stark et al.\ 2009; Gonzalez et al.\ 2010;
McLure et al.\ 2011) largely due to the presence of strong nebular
emission lines (Shim et al.\ 2011; Schaerer \& de Barros 2010; Wilkins
et al. 2013b), but now seems clear from $z\sim6$ to $z\sim2$ (Stark et
al.\ 2013; Gonzalez et al.\ 2014; Schaerer \& de Barros 2013;
Labb{\'e} et al.\ 2013; Smit et al.\ 2014; Salmon et al.\ 2015).

The spectral characteristics of galaxies at $z\sim8$ and $z\sim8.5$
have also been explored (Bouwens et al.\ 2010, 2013; Finkelstein et
al.\ 2010, 2012; Dunlop et al.\ 2013), but are more difficult to
robustly quantify due to the limited leverage in wavelength available
for constraining these slopes and corrections required to remove the
impact of the IGM absoprtion (Bouwens et al.\ 2014a).  This is
particularly true for $UV$-continuum slope determinations at
$z\sim8.5$-9.5.  Given the challenges in deriving the spectral
characteristics of galaxies at $z\sim8.5$-9.5, it may seem that
further advances may need to wait until the {\em James Webb Space
  Telescope} (JWST).

Fortunately, as we will show, we can make immediate progress on this
issue by taking advantage of the deep IRAC observations available for
galaxy samples at $z\sim10$ from the {\em Spitzer Space Telescope}.
Over the redshift interval $z\sim9.5$-10.5, the {\em Hubble} $H$-band
band and the IRAC $3.6\mu$m band fall in the $UV$-continuum.  The
wavelength leverage is sufficient between these bands that one can
plausibly quantify the UV-continuum slopes of galaxies more accurately
at $z\sim10$ than is possible at $z\sim8$ and especially at
$z\sim8.5$-9, as we demonstrate in \S\ref{sec:observations.slope} of
this paper.

Opportunities to use $z\sim10$ samples to perform such studies now
exist thanks to the deep near-IR data available from {\em Hubble} in
the Cosmic Assembly Near-Infrared Deep Extragalactic Survey (CANDELS,
Grogin et al. 2011; Koekemoer et al. 2011), {\em Hubble} Ultra-Deep
Field 2009/2012 programs (Bouwens et al.\ 2011; Koekemoer et al. 2013;
Ellis et al. 2013)\footnote{The HUDF12 programme \ obtained
  $Y_{f105w}$ and $J_{f160w}$. Both the HUDF09 and HUDF12
  observations, along with imaging from various other sources have
  been released as part of the eXtreme Deep Field (XDF) project
  (Illingworth et al. 2013).}, and Cluster Lensing and Supernovae
survey with {\em Hubble} (CLASH: Postman et al.\ 2012) programs.  The
samples range from particularly faint sources over the HUDF/XDF (Ellis
et al. 2013; Oesch et al. 2013) and the Frontier Fields (Zitrin et
al.\ 2014) to brighter sources located over CANDELS (Oesch et
al. 2014) and CLASH (e.g. Zheng et al. 2012, Coe et al. 2013, and
\ Zitrin et al. 2014).\footnote{It is important to note that not all
  such candidates necessarily lie at $z\sim 10$; indeed UDFj-39546284
  (Bouwens et al. 2011a) for instance, which was initially reported to
  be a $z\sim 10$ candidate, no longer has a favoured high-redshift
  identification (Ellis et al.\ 2013; Brammer et al. 2013, Bouwens et
  al. 2013a).}

Particularly valuable for probing the spectral characteristics of
galaxies in the early universe are those $z\sim10$ candidates that are
intrinsically bright or lensed, since those sources have sufficient
S/N with IRAC that one can use them to quantify the mean
$UV$-continuum slope of galaxies at very early times.  To date, five
such galaxies have been identified in the magnitude range 26-27 mag
from the CANDELS fields (Oesch et al.\ 2014) and behind lensing
clusters (Zheng et al.\ 2012).

In this paper, we make use of these 5 particularly bright $z\sim10$
candidates to derive, for the first time, the mean $UV$-continuum
slope $\beta$ at $z\sim10$ for a multi-object sample.  We compare the
observed properties with predictions from recent cosmological galaxy
formation simulations to provide some context.  Such simulations are
valuable given that the $UV$-continuum slopes and $UV$-optical colours
can be affected by a complex mixture of different factors including
the joint distribution of stellar masses, ages, metallicities, and the
escape fraction -- which makes it difficult to interpret the
observations in terms of a single variable (Bouwens et al.\ 2010;
Wilkins et al.\ 2011).

This paper is organised as follows: in Section \ref{sec:observations}
we describe recent observations of candidate $z\sim 10$ star forming
galaxies. In Section \ref{sec:dust_interp} we interpret observations
of these systems in the context of dust emission.  Finally, in Section
\ref{sec:conclusions} we present our conclusions. Throughout this work
magnitudes are calculated using the $AB$ system (Oke \& Gunn 1983). In
calculating absolute magnitudes we assume a $\Omega_{M}=0.3$,
$\Omega_{\Lambda}=0.7$, $h=0.7$ cosmology.

\section{Observations of the UV-continuum slope}\label{sec:observations}

\subsection{Data and Sample Selection}\label{sec:data}

This paper is based on the bright $z\sim10$ galaxy sample selected
over the GOODS fields from Oesch et al. (2014) as well the $z\sim9.6$
CLASH source behind MACS1149 identified in Zheng et al. (2012).  For
details on the datasets, we refer the reader to the discovery
papers. We summarise the photometry of these sources in Table
\ref{tab:photometry}.

In brief, the candidates from Oesch et al. (2014) were identified
using the complete {\em Hubble} dataset from the CANDELS survey
(Grogin et al. 2011, Koekemoer et al. 2011) in addition to ancillary
Advanced Camera for Surveys (ACS) data mostly from the GOODS
survey. The central area of the GOODS-South and North fields represent
the CANDELS Deep survey, which reach to H$_{f160w}=27.8$ mag
($5\sigma$), while the outer regions are covered with slightly
shallower data H$_{f160w}=27.1$ mag.  Galaxy candidates at $z\sim10$
were identified based on the spectral break shortward of
Lyman-$\alpha$ resulting in a red color of
$($J$_{f125w}-$H$_{f160w})>1.2$ and complete non-detection in all
shorter wavelength filters. While Oesch et al. (2014) also selected
$z\sim9$ galaxies with a weaker continuum break, here we restrict our
analysis to the redder color selection resulting in galaxies with
$z_{phot}\gtrsim9.5$.

Here we make use of a redetermination of the IRAC photometry for the
four sources in the Oesch et al.\ (2014) sample.  For these
measurements, we take advantage of new reductions of the Spitzer/IRAC
3.6\,$\mu$m and 4.5\,$\mu$m imaging data over the GOODS fields from
Labbe et al.\ (2015).  These reductions include all data from the
original GOODS, the Spitzer Extended Deep Survey (SEDS: Ashby et
al.\ 2013), the IRAC Ultra Deep Field (IUDF: Labb\'e et al.\ 2015),
and S-CANDELS (Ashby et al.\ 2015) programs.  The average 5$\sigma$
depths of the IRAC data within 1\arcsec\ radius apertures are 27.0 and
26.7 mag in the two channels, respectively.

Our measurements again make use of the \textsc{mophongo} software
(Labb{\'e} et al. 2006, 2010, 2013) to model and subtract the flux
from neighboring sources, but include significantly improved PSF
modeling.  Our derived PSFs account for the orientation of all
exposures that contribute to our IRAC reductions.  Our rederived IRAC
fluxes are completely consistent within the errors with those given in
Oesch et al. (2014).  Three of the four $z\sim10$ galaxy candidates
from Oesch et al. (2014) are significantly detected ($>4.5\sigma$) in
these data in at least one filter. In particular, the brightest source
GN-z10-1 with a photometric redshift of $z_{phot} = 10.2\pm0.4$ is
robustly detected in the 3.6\,$\mu$m channel at $\sim7\sigma$,
allowing for an individual estimate a UV-continuum slope for this
source.

Equally important for robust measurements of the $H_{160}-[3.6]$
colors are accurate measurements of the total $H_{160}$-band flux.
Our determination of the total $H_{160}$-band flux is as described in
Oesch et al.\ (2014) and includes all the light inside an elliptical
aperture extending to 2.5 Kron (1980) radii.  The measured flux inside
the utilized Kron aperture is corrected to total (typically a
$\sim$0.2 mag correction) based on the expected light outside this
aperture (Dressel et al.\ 2012).

As a correction to total is performed for both the HST $H_{160}$-band
photometry and for the Spitzer/IRAC photometry, colors derived from
HST to Spitzer/IRAC should not suffer from significant biases.
Nevertheless, it was worthwhile to verify that this was the case by
applying the HST-to-Spitzer PSF-correction kernel to the HST data
(derived from \textsc{mophongo}) and then measuring total magnitudes
from the HST data in the same way as the Spitzer/IRAC data.  Given the
relatively small number of $z\sim10$ candidates in our sample and
their limited S/N after convolving with the IRAC PSF correction
kernel, we perform this test on a sample of bright ($H_{160,AB}<25$)
$z\sim4$ galaxies from the Bouwens et al.\ (2015a) catalogs.  We found
that the total magnitudes we recover by applying this procedure to the
HST observations of these $z\sim4$ sources were consistent to $<0.1$
mag in the median with that derived using our primary method.

To probe intrinsically fainter galaxies while maintaining a sufficiently high signal-to-noise, we can take advantage of sources which are lensed by foreground galaxies (or clusters of galaxies). Several $z\sim 10$ candidates have now been discovered in cluster searches (Zheng et al. 2012; Coe et al. 2013; Zitrin et al. 2014), with more likely to be identified in the near future as a result of the ongoing {\em Hubble} Frontier Fields observations. 

Of the three currently known $z\sim 10$ candidates we concentrate on MACS1149-JD (Zheng et al. 2012). The object presented by Zitrin et al. (2014) does not currently have sufficiently deep IRAC imaging to provide anything other than a weak upper limit on the rest-frame UV-continuum slope. The object presented by Coe et al. (2013) has a photometric redshift of $z\sim 10.7$, at which redshift the $H_{160}$-band flux could be affected by the position of the Lyman-$\alpha$ break within the $H_{160}$ band and/or Lyman-$\alpha$ line emission. 

MACS1149-JD was identified in a search of 12 CLASH clusters and has strong detections in both the JH$_{f140w}$ and H$_{f160w}$ bands with weaker detections in both Y$_{f105w}$ and J$_{f125w}$ and non-detections in several optical bands. Photometric redshift fitting of the sources photometry suggests it is $z\sim 9.6$.\footnote{Bouwens et al.\ 2014b estimate a photometric redshift of 9.7$\pm$0.1.}  Zheng et al. (2012) presented {\em Spitzer}/IRAC photometry of MACS1149-JD ([3.6]$<160\,{\rm nJy},\,1\sigma$) based on observations taken under Program ID 60034 (PI: Egami). We augment these observations with Frontier Fields observations and archival data taken as part of the Spitzer UltRa Faint SUrvey Program (Surfs'Up, Brada{\v c} et al. 2014) and measure a [3.6] flux of $175\pm44$ nJy (see Bouwens et al.\ 2014b). This is consistent with that reported by both Zheng et al. (2012) ($<160\,{\rm nJy},\,1\sigma$) and Brada{\v c} et al. (2014) ($190\pm 87\,{\rm nJy}$).

\begin{table*}
\caption{Photometry and derived properties for the $z\sim10$ sources, and the bright stack, considered in this work.\label{tab:photometry}}
\begin{tabular}{lccccccccc}
\hline
 & $z$ & $M_{1500}$ & H$_{f160w}$/nJy & [3.6]/nJy & H$_{f160w}-$[3.6] & $\beta_{\rm obs}$ & $A_{1500,{\rm C00}}$ & $A_{1500,{\rm SMC}}$ &\\
& & & & & & & \multicolumn{2}{c}{(assumes $\beta_{\rm int} = -2.54$)$^7$} \\
\hline
GN-z10-1$^{1,2}$ & $10.2$ & $-21.6$ & $152\pm10$ & $136\pm29$ & $-0.1\pm0.2$ & $-2.1\pm0.3$ & $0.9\pm0.6$ & $0.5\pm0.3$ \\
GN-z10-2$^{1,2}$ & $9.8$ & $-20.7$ & $68\pm9$ & $45\pm26$ & $-0.5\pm0.6$ & $-2.5\pm0.7$ & $0.1\pm1.6$ & $0.1\pm0.9$ \\
GN-z10-3$^{1,2}$ & $9.5$ & $-20.6$ & $73\pm8$ & $59\pm24$ & $-0.2\pm0.5$ & $-2.3\pm0.5$ & $0.6\pm1.1$ & $0.3\pm0.6$ \\
GS-z10-1$^{1,2}$ & $9.9$ & $-20.6$ & $66\pm9$ & $71\pm27$ & $0.1\pm0.4$ & $-1.9\pm0.5$ & $1.4\pm1.1$ & $0.8\pm0.6$ \\
MACS1149-JD1$^{4}$ & $9.6$ & $-19.4$$^4$ & $190\pm13.3$$^5$ & $177\pm44$ & $-0.1\pm0.3$ & $-2.1\pm0.3$ & $1.0\pm0.7$ & $0.6\pm0.4$ \\
\hline
\bf STACK & - & - &  - &  - & $-0.1\pm0.2$$^6$ & $-2.1\pm0.3$$^6$ & $0.9\pm0.6$ & $0.5\pm0.3$  \\
\hline
\end{tabular}

\medskip
$^1$ Oesch et al. (2014). $^{2}$ included in stack. $^3$ Zheng et al. (2012), Bouwens et al.\ (2014b). $^{4}$ MACS1149-JD is gravitationally lensed by a foreground cluster, we determine the un-lensed absolute magnitude using the best-fit magnification of $14.5$. $^{5}$ We independently measure the [3.6] flux for MACS1149-JD using the Frontier Fields {\em Spitzer}/IRAC observations combined with observations taken as part of the Spitzer UltRa Faint SUrvey Program (SURFSUP, Brada{\v c} et al. 2014) and Program ID 60034 (PI: Egami). Our flux measurements are consistent with those reported by both Zheng et al. (2012) ($<160\,{\rm nJy},\,1\sigma$) and Brada{\v c} et al. (2014) ($190\pm 87\,{\rm nJy}$).  $^6$ The uncertainty on the mean increases to $\pm0.3$, if we assume there is significant intrinsic scatter in the $\beta$ distribution as observed at $z\sim4$-5 (Bouwens et al.\ 2009, 2012; Castellano et al.\ 2012; Rogers et al.\ 2014: see \S2.2).  $^7$ The intrinsic $UV$-continuum slope predicted in dynamical simulation at $z\sim 10$ assuming $f_{\rm esc}=0$ (\S3).
\end{table*}

\subsubsection{Bright Stack}\label{sec:bright_stack}

The uncertainties on the $UV$-continuum slopes of individual sources
is large enough that it is useful to consider constraints for the
average $z\sim10$ source from the Oesch et al.\ (2014) bright sample.
We therefore stack the photometry of the 4 $z\sim 10$ candidates
presented in Oesch et al. (2014).  The H$_{f160w}-$[3.6] colour of the
stack is $-0.1\pm0.2$. If we stack only GN-z10-1 and GN-z10-2 (i.e. excluding sources without JH$_{f140w}$ detections) we find H$_{f160w}-$[3.6]$=-0.2\pm0.2$.

\subsection{Measuring the UV continuum slope}\label{sec:observations.slope}

\begin{figure}
\centering
\includegraphics[width=20pc]{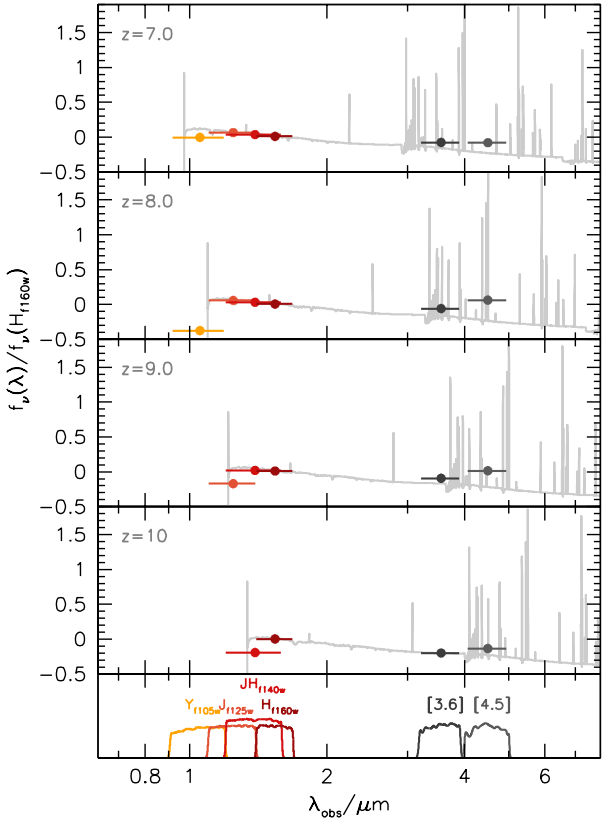}
\caption{Relative observed near-IR photometry of a model star-forming galaxy at $z\in\{7,8,9,10\}$ highlighting the bands available to measure the rest-frame UV-continuum slope. At $z>9.6$, the {\em Spitzer}/IRAC [3.6] band can be combined with the H$_{f160w}$ band to measure the UV continuum slope over a large wavelength baseline, minimising its uncertainty. At $z\sim 8$ only the JH$_{f140w}$ and H$_{f160w}$ bands are uncontaminated by the Lyman-$\gamma$ break providing only a small wavelength baseline and leaving the uncertainty on the observed UV continuum slope very large.}
\label{fig:measuring_beta}
\end{figure}

\begin{figure}
\centering
\includegraphics[width=20pc]{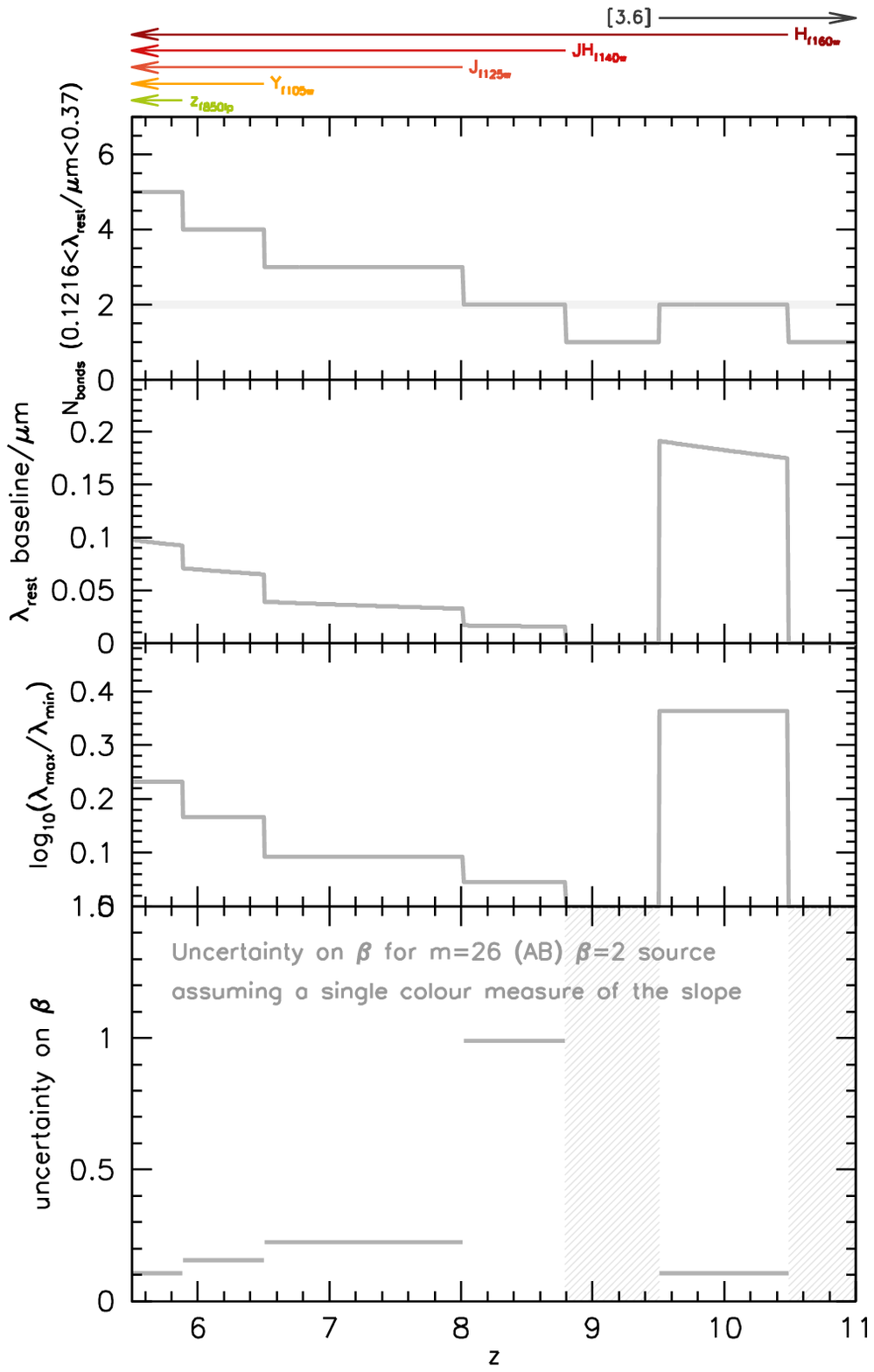}
\caption{{\em Top panel} - The number of {\em Hubble}/WFC3 (Y$_{f098m}$, J$_{f125w}$, JH$_{f140w}$, and H$_{f160w}$) and {\em Spitzer}/IRAC ([3.6]) bands probing the rest frame $1216<\lambda_{\rm rest}/{\rm\AA}<3700$ UV continuum as a function of redshift. {\em Second panel} - The rest-frame UV wavelength baseline accessible by {\em Hubble}/WFC3 and {\em Spitzer}/IRAC observations as a function of redshift. {\em Third panel } - The ratio of the mean wavelength of the bluest and reddest filters to probe the rest-frame UV continuum. The arrows at the top denote the redshift range over which an individual band probes the rest-frame $1216<\lambda_{\rm rest}/{\rm\AA}<3700$ UV continuum. {\em Bottom panel } - The expected uncertainty on the measurement of $\beta$ as a function of redshift. This assumes a $\beta=-2$, $m=26$ source, for which $\beta$ is measured from the colour providing the longest usable wavelength baseline.}
\label{fig:wavelength_baseline}
\end{figure}

For $z>7$ galaxies, observations of the rest-frame UV continuum are
typically limited to just 3 WFC3/IR bands uncontaminated by the
Lyman-$\alpha$ break (i.e., $J_{f125w}$, $JH_{f140w}$, and
$H_{f160w}$), and therefore suitable to measure the UV continuum slope
$\beta$. This is demonstrated, in Figure \ref{fig:measuring_beta}, for
sources at $z\in\{7,8,9,10\}$.

When only two bands are utilised for $UV$-continuum slope estimates, the relationship\footnote{Assuming the underlying spectral flux density is described by a power-law, i.e. $f_{\nu}\propto\lambda^{\beta+2}$} between the observed colour and UV continuum slope can simply be written as:

\begin{equation}
\beta = p\times (m_{1}-m_{2})_{\rm AB}-2,
\end{equation}
where $(m_{1}-m_{2})_{\rm AB}$ is the observed colour assuming the AB magnitude system and $p$ is a value sensitive to the choice of bands. The value $p$ is approximately related to the ratio of the effective wavelengths ($\lambda_1$, $\lambda_2$) of the filters used to probe the slope\footnote{The value of $p$ is also sensitive to the shape of the filter and when calculating the final value appropriate for our observations we take this into account.}, i.e.,

\begin{equation}
p^{-1} = 2.5\cdot\log_{10}(\lambda_1/\lambda_2).
\end{equation}

The value of $\log_{10}(\lambda_1/\lambda_2)$ and the rest-frame UV wavelength baseline ($\lambda_2-\lambda_1$) accessible by {\em Hubble}/WFC3 and {\em Spitzer}/IRAC observations as a function of redshift are shown in Figure \ref{fig:wavelength_baseline}.

This accessible wavelength baseline is important as the uncertainty on $\beta$ will also scale with $p$. For example, at $z\sim 8$ where we only have observations using the JH$_{f140w}$ and H$_{f160w}$ bands, the value of $p$ using these bands is approximately $9.0$ (Bouwens et al.\ 2014a). A typical error on $m_{1}$ and $m_{2}$ of $0.1$ then translates into a large uncertainty on $\beta$ ($\approx 9.0\cdot\sqrt{0.1^{2}+0.1^{2}}\approx 1.3$). At $z>9.5$ the {\em Spitzer}/IRAC [3.6]-band probes the UV continuum, which, when combined with H$_{f160w}$-band observations, provides an especially extended wavelength baseline.  The end result is a particularly small value for $p$ ($\approx 1.1$) and thus small uncertainty on $\beta$. This long wavelength baseline compensates for the lower sensitivity of observations with {\em Spitzer}/IRAC [3.6], allowing the the UV continuum to be estimated much more robustly than at $z\sim 8$ and on a par with $z\sim 6-8$ for the same observed apparent magnitude. This is demonstrated in the bottom panel of Figure \ref{fig:wavelength_baseline}. 

\begin{figure}
\centering
\includegraphics[width=20pc]{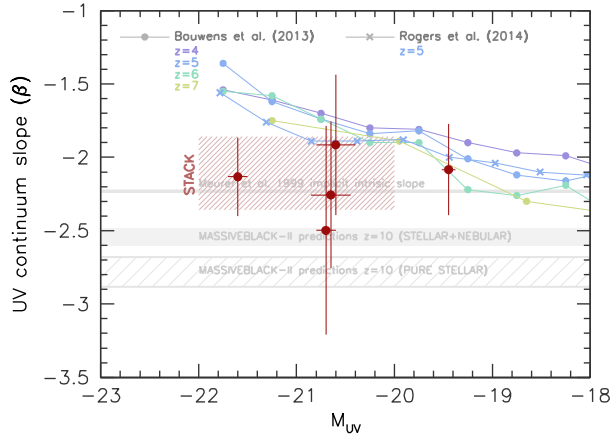}
\caption{Mean value of the UV-continuum slope as a function of absolute rest-frame UV magnitude for candidate high-redshift ($z=4-10$) star forming galaxies (Bouwens et al.\ 2014; Rogers et al.\ 2014). The thin horizontal line denotes the intrinsic slope implicit in the empirical Meurer et al. (1999) relation and the two horizontal bands show the range of intrinsic slopes expected from the {\sc MassiveBlack-II} hydrodynamical simulation at $z\sim 10$ assuming both $f_{\rm esc}=1$ (pure stellar) and $f_{\rm esc}=0$. It is important to note that at lower redshift the {\sc MassiveBlack-II}  simulation predicts significantly bluer intrinsic slopes (see Wilkins et al. 2013a).}
\label{fig:M_beta}
\end{figure}

The observed values of the UV-continuum slope $\beta$ for the $z\sim
10$ candidates (and the stack) are listed in Table
\ref{tab:photometry} and shown in Figure \ref{fig:M_beta}. For the
brightest candidate (GN-z10-1) we find $\beta_{\rm obs}=-2.1\pm 0.2$
while for the bright stack (see \S\ref{sec:bright_stack}) we find
$\beta_{\rm obs}=-2.1\pm 0.3$.  If we stack only those sources which
have $JH_{140}$-band observations (providing a second WFC3/IR filter
where the $z\sim10$ candidates are detected), i.e., GN-z10-1 and
GN-z10-2, we find $\beta_{\rm obs}=-2.2\pm 0.3$.

While the formal random error on the mean $\beta$ is 0.2, at lower
redshifts the $\beta$ distribution for luminous galaxies appears to
show a significant intrinsic scatter of $\sigma_{\beta}\sim0.35$
(Bouwens et al.\ 2009, 2012; Castellano et al.\ 2012; Rogers et
al.\ 2014).  Assuming a similar scatter at $z\sim10$ translates to a
slightly larger random error on the mean $\beta$ of 0.3.

\subsubsection{Selection Biases}

It is useful to consider briefly whether our mean $\beta$ results
could be biased because of the selection criteria that were applied in
searching for $z\sim10$ galaxies.  Such issues became an important
aspect of the debate regarding $UV$-continuum slopes at $z\sim7$
(e.g., Wilkins et al.\ 2012; Dunlop et al.\ 2012; Bouwens et
al.\ 2012, 2014a), and it is important that we ensure that such issues
do not become important again.


In identifying bright $z\sim9$-10 candidates over CANDELS GOODS-S and
GOODS-N, Oesch et al. (2014) did not consider sources with
particularly red H$_{f160w}-$[4.5]$>2$ colors redward of the Lyman
break.  Although the H$_{f160w}-$[4.5] colour, unlike the
H$_{f160w}-$[3.6] colour, does not directly probe the UV continuum
slope, they are correlated and noise in the two colors is not
independent.\footnote{This is particularly true at
  H$_{f160w}-$[4.5]$>0$ where the colour is dominated by the effect of
  dust reddening.}. A colour of H$_{f160w}-[4.5]\sim2$ corresponds to
$\beta\approx -0.1$.  Given that all of our candidates have measured
colors of $\beta<-1.8$ with small uncertainties, it is unlikely that a
$\beta<-0.1$ selection will have an important impact on the mean
$\beta$ measured for the sample, since the limit is $>3\sigma$ away
from the mean $\beta$ observed and $3\sigma$ away from the mean
$\beta$ assuming no evolution from $z\sim5$-7.

It is encouraging that Bouwens et al.\ (2015a) selected exactly the
same $z\sim10$ candidates as Oesch et al.\ (2014), utilizing a
slightly modified criteria (H$_{f160w}-$[3.6]$<1.6$: corresponding to
$\beta<-0.2$).  This further confirms that the Oesch et al. (2014)
sample of bright sources is not affected by strong $\beta$-dependent
selection biases.

\subsection{Possible Evolution in the observed UV-continuum slope}

In the previous section, we presented the first determination of the
mean $UV$-continuum slope $\beta$ for a multi-object sample at
$z\sim10$.  Previously, UV continuum slopes for similarly luminous
galaxies could only be determined up to $z\sim8$ (Finkelstein et
al.\ 2012).

\begin{figure}
\centering
\includegraphics[width=20pc]{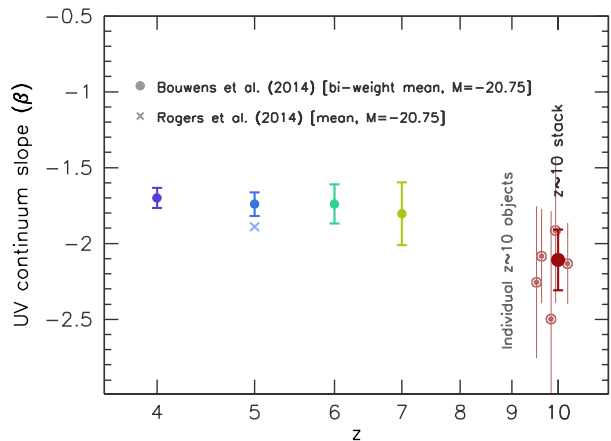}
\caption{The observed UV continuum slope as a function of redshift for candidate high-redshift ($z=4-10$) star forming galaxies. Results at $z=4-7$ are based on the bi-weight mean of galaxies with $M_{1500}\approx -20.75$ from Bouwens et al. (2014a).}
\label{fig:beta_z_M}
\end{figure}

With these new measurements in hand, it is interesting to look for
evidence of evolution in the mean $\beta$ of galaxies versus redshift.
Figure~\ref{fig:beta_z_M} presents the current measurements and
compares it against previous measurements at $z\sim4$-7 from Bouwens
et al.\ (2014a) and Rogers et al.\ (2014).

As is evident from Figure~\ref{fig:beta_z_M}, the observed UV
continuum slopes of individual galaxies at $z\sim 10$ are found to be
much bluer ($\Delta\beta\approx 0.4$) than those at $z\sim 4$-7.
These new results are interesting, as they suggest either a gradual
evolution in $\beta$ towards bluer colors or no evolution over a wide
redshift baseline.

The comparison of measurements at $z=4$-7 with $z\sim 10$ is not
necessarily straightforward as the $UV$-continuum slope is measured
over very different wavelength baselines which may introduce a
systematic bias.  This is explored in more detail in Appendix A where
we conclude that this is unlikely to be an important factor in this
case, but could be as large as $\Delta\beta\sim 0.2$.  If we allow for
potential $\sim$10\% systematics in our HST - IRAC color measurements
(possible if the total magnitudes we estimate from the HST or
Spitzer/IRAC data are not quite identical), the total systematic error
relevant to the inferred evolution in $\beta$ could be as large as
$\sim0.22$.

Accounting for both random and systematic errors, we find the observed
$\beta$ for luminous galaxies at $z\sim10$ is only $1\sigma$ bluer
than at $z\sim7$.  Therefore, our results are consistent with either a
mild reddening of the $UV$-continuum slopes with cosmic time or no
evolution at all.

\section{Inferred Dust Attenuation}\label{sec:dust_interp}


\subsection{Relating the observed UV continuum slope to dust attenuation}

The presence of dust causes the observed UV continuum slope to redden relative to the intrinsic slope. The relationship between between the observed slope $\beta_{\,\rm obs}$ and the attenuation $A_{\lambda}$ can be written as (e.g. Meurer et al. 1999, Wilkins et al. 2013a),
\begin{equation}\label{eq:abeta}
A_{\lambda}=D_{\lambda}\times[\beta_{\,\rm obs}-\beta_{\,\rm int}],
\end{equation} 
where $\beta_{\rm int}$ is the intrinsic UV continuum slope, $\beta_{\rm obs}$ is the observed slope, and $D_{\lambda}$ ($={\rm d}A_{\lambda}/{\rm d}\beta$) describes the change in the attenuation as a function of the change in $\beta$, and is sensitive to the choice of attenuation/extinction curve. The intrinsic UV continuum slope is sensitive to a range of properties, including the star formation and metal enrichment histories, and the ionising photon escape fraction (see \S\ref{sec:dust_interp.intbeta}, and Wilkins et al. 2012, Wilkins et al. 2013a). 

Both $\beta_{\rm int}$ and $D_{\lambda}$ can be constrained empirically (e.g. Meurer et al. 1999, Heinis et al. 2013) using a combination of rest-frame UV and far-IR observations of a sample of galaxies. $D_{\lambda}$ can be determined for any attenuation curve which extends over the rest-frame UV, and thus doesn't necessarily require far-IR observations to be constrained. 

\subsection{The intrinsic UV continuum slope}\label{sec:dust_interp.intbeta}

Empirical constraints on $\beta_{\rm int}$ are, at present, due to the lack of sufficiently deep far-IR/sub-mm observations, limited to low-intermediate redshift (e.g. Heinis et al. 2013). The value of $\beta_{\rm int}$ at low/intermediate redshift is unlikely to reflect that at very-high redshift as $\beta_{\rm int}$ is sensitive to both the age and metallicity of the stellar population, both of which are expected to decrease in typical star forming galaxies to high-redshift (e.g. Wilkins et al. 2013a).

\begin{figure}
\centering
\includegraphics[width=20pc]{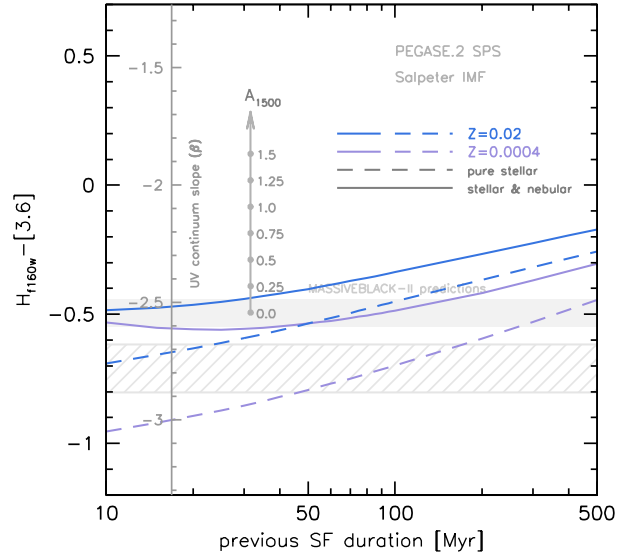}
\caption{The sensitivity of the H$_{f160w}$-[3.6] colour (our proxy for the UV continuum slope) to the duration of previous (constant) star formation. The blue and purple lines show the result for $Z=0.02$ and $Z=0.0004$ respectively, while the solid and dashed lines show the result assuming $f_{\rm esc}=0$ (i.e. including nebular continuum and line emission) and $f_{\rm esc}=1$ (i.e. pure stellar emission) respectively. The solid and hatched horizontal bands show the predictions from the {\sc MassiveBlack-II} simulations assuming $f_{\rm esc}=0$ and $f_{\rm esc}=1$ respectively.}
\label{fig:colours_ageZ_constant_z10}
\end{figure}

The sensitivity of the intrinsic UV-continuum slope $\beta_{\rm int}$ to various properties, including the joint distribution of stellar masses, ages, and metallicities (themselves determined the recent star formation and metal enrichment histories, and initial mass function) and the presence of nebular continuum, and to a lesser extent, line emission is discussed in Wilkins et al. (2012) and Wilkins et al. (2013a). We demonstrate this sensitivity in Figure \ref{fig:colours_ageZ_constant_z10} utilising the {\sc Pegase.2} stellar population synthesis code (Fioc \& Rocca-Volmerange 1997, 1999). We determine the H$_{f160w}-$[3.6] colour (our proxy for the UV continuum slope at $z\sim 10$) as a function of the duration of previous constant star formation, for two stellar metallicities, and also assuming $f_{\rm esc}=0$ and $f_{\rm esc}=1$ (i.e. a pure stellar continuum). Increasing the metallicity, the strength of nebular emission (i.e. decreasing the escape fraction $f_{\rm esc}$), and the duration of previous star formation all result in the intrinsic UV continuum slope becoming redder. 

\subsubsection{Sensitivity to the candidate redshift}

\begin{figure}
\centering
\includegraphics[width=20pc]{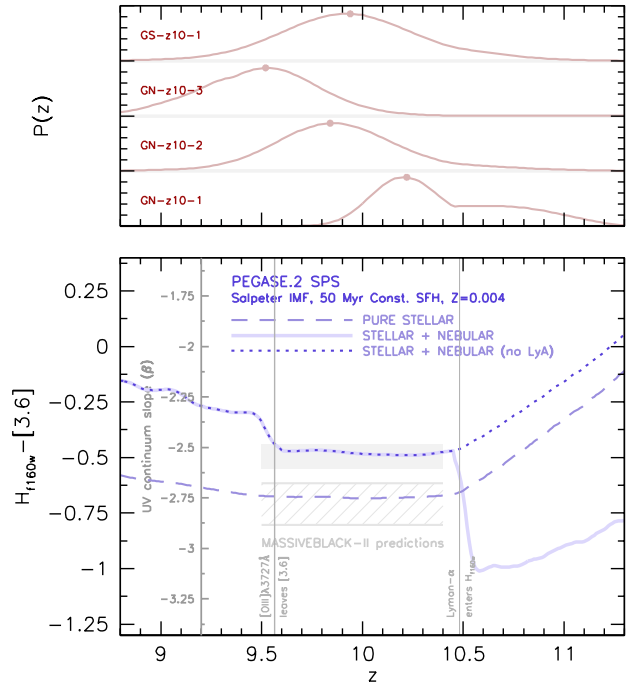}
\caption{{\em lower-panel} The sensitivity of the H$_{f160w}$-[3.6] colour (our proxy for the UV continuum slope) to the redshift for a stellar population which has been constantly forming stars for 50 Myr (with $Z_{}=0.004$, i.e. $\approx 1/5\,Z_{\odot}$). The three lines show the result of different nebular emission scenarios: ({\em dashed}) $f_{\rm esc}=1$ (i.e. pure stellar), ({\em solid}) $f_{\rm esc}=0$, and ({\em dotted}) $f_{\rm esc}=1$ but with Lyman-$\alpha$ suppressed. The predictions from {\sc MassiveBlack-II} are shown by the shaded horizontal bands. {\em upper-panel} The redshift probability distributions of the 4 bright $z\sim 10$ candidates.}
\label{fig:colours_z}
\end{figure}

In addition to the physical properties outlined above the UV continuum slope inferred from observations will also be sensitive to the redshift of the source. This is demonstrated in Figure \ref{fig:colours_z} where the predicted intrinsic H$_{f160w}-$[3.6] colour (our proxy for the observed UV-continuum slope) of a stellar population that has been forming stars constantly for 50 Myr is shown as a function of redshift. Three different nebular emission scenarios are shown: (i) $f_{\rm esc}=1$ (i.e. pure stellar), (ii) $f_{\rm esc}=0$, and (iii) $f_{\rm esc}=1$ but with Lyman-$\alpha$ suppressed. In all three cases, in the interval $z=9.6-10.4$, the observed colour exhibits virtually no variation. At $z<9.6$ various strong emission lines, beginning with [OII]$\lambda 3727{\rm\AA}$, enter the {\em Spitzer}/IRAC [3.6]-band resulting in generally redder colours\footnote{The {\em Spitzer}/IRAC [3.6]-band also no longer probes the rest-frame UV continuum}.

At $z>10.4$ the H$_{f160w}$-band filter overlaps with the wavelength
of rest-frame Lyman-$\alpha$. Assuming no Lyman-$\alpha$ emerges
(scenario {\em iii}) this results in the observed H$_{f160w}$-band
encompassing the Lyman-$\alpha$ break resulting in a decreased flux,
and consequently redder H$_{f160w}-$[3.6] colours. If strong
Lyman-$\alpha$ emerges the H$_{f160w}-$[3.6] colour would rapidly
become very blue before gradually becoming redder.  Spectroscopic
follow-up of $z\gtrsim6.5$ galaxies strongly points towards very
little Ly$\alpha$ emission in $z\gtrsim10.4$ galaxies (e.g., Stark et
al.\ 2010; Ono et al.\ 2012; Schenker et al.\ 2012; Pentericci et
al.\ 2011; Caruana et al.\ 2012; Treu et al.\ 2013; Finkelstein et
al.\ 2013; Oesch et al.\ 2015a: but see also Zitrin et al.\ 2015).

\subsubsection{Simulation Predictions for the Intrinsic Slope}

While empirical constraints on the intrinsic slope $\beta_{\rm int}$ exist they are only available at low-redshift, and are therefore unlikely to be representative of the very-high redshift Universe. We then also employ predictions from galaxy formation models for the intrinsic slope and spectral energy distribution. Specifically, we utilise predictions from the {\sc MassiveBlack} and {\sc MassiveBlack-II} hydro-dynamical simulations (see Khandai et al. 2015 for a general description of the simulations, and Wilkins et al. 2013a for predictions of the intrinsic UV continuum slope). At $z=10$ these simulations predict a median intrinsic slope of $\beta_{\rm int}\approx -2.54$ (assuming $f_{\rm esc}=0$) and $\beta_{\rm int}\approx -2.78$ for a pure stellar SED. In both cases the {\sc Pegase.2} SPS model is assumed, however other commonly used models models produce similar results (Wilkins et al. 2013ab).

\subsection{Inferred Dust Attenuation}

By combining the observed UV continuum slope $\beta_{\rm obs}$ with a choice of intrinsic slope $\beta_{\rm int}$ and dust curve we can infer the level of dust attenuation using Equation \ref{eq:abeta}. We do this both utilising the empirical Meurer relation and by assuming the intrinsic slope predicted by the {\sc MassiveBlack-II} simulation combined with various attenuation/extinction curves. 

\begin{figure}
\centering
\includegraphics[width=20pc]{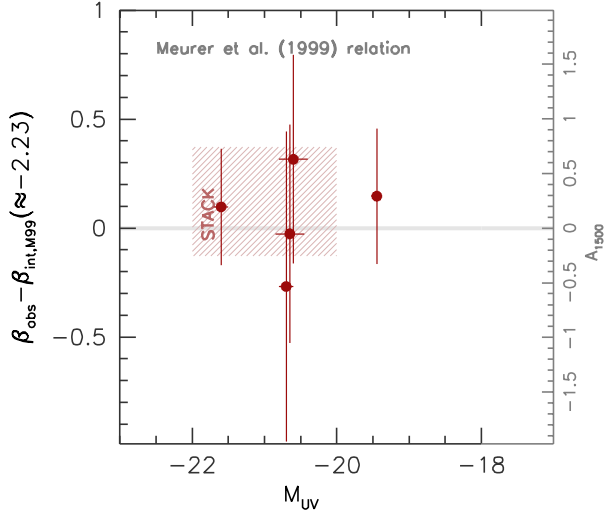}
\caption{The inferred rest-frame UV attenuation $A_{1500}$ (right-hand axes) assuming the Meurer et al. (1999) relation.}
\label{fig:M_A1500_M99}
\end{figure}

In Figure \ref{fig:M_A1500_M99} we first show the inferred UV
attenuation assuming the empirical Meurer et al. (1999) relation. The
combination of very blue observed colours and the $\beta\approx -2.23$
intrinsic slope implicit in the Meurer relation result in individual objects
having best fit attenuations formally consistent with $A_{1500}=0$. The attenuation inferred from the bright stack is $A_{1500}=0.3\pm 0.5$.

\begin{figure}
\centering
\includegraphics[width=20pc]{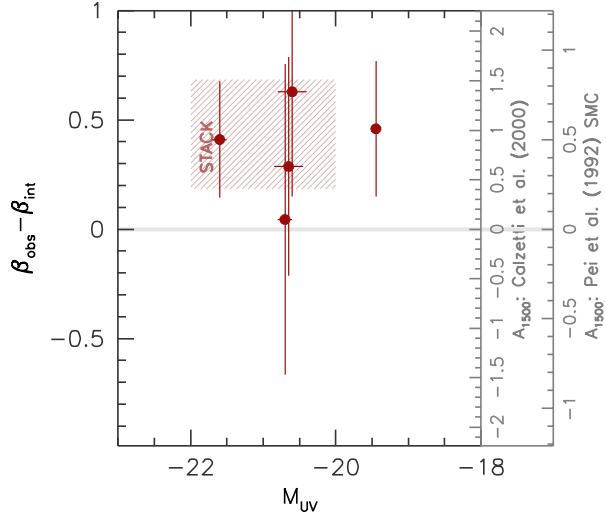}
\caption{The inferred rest-frame UV attenuation $A_{1500}$ (right-hand axes) assuming the intrinsic slope predicted by {\sc MassiveBlack} along with the Calzetti et al. (2000) {\em starburst} and Pei et al. (1992) SMC dust curves.}
\label{fig:M_A1500}
\end{figure}

Figure \ref{fig:M_A1500} is similar but instead shows the attenuation when we assume the intrinsic slope predicted by the {\sc MassiveBlack-II} simulation (assuming $f_{\rm esc}=0.0$) along with both the Calzetti et al. (2000) and SMC (from Pei et al. 1992) dust curves. In this case all of the individual observations as well of the stack yield positive values of $A_{1500}$. If instead a higher escape fraction is assumed (which would make the intrinsic slope bluer) the inferred attenuation increases by between $0.25-0.5$ mags, depending on the choice of attenuation/extinction curve.

\subsubsection{Evolution of dust attenuation}

\begin{figure}
\centering
\includegraphics[width=20pc]{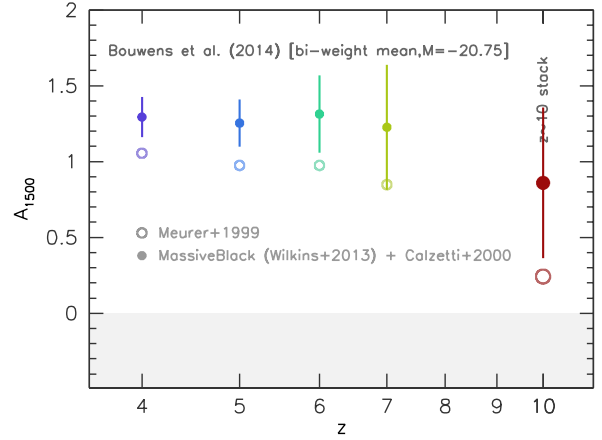}
\caption{The evolution of the rest-frame UV attenuation $A_{1500}$ inferred from observations of the UV continuum slope in bright $z\in\{4,5,6,7,10\}$ star forming galaxies. Both the attenuation calculated using the Meurer et al. (1999) relation and combining the Calzetti et al. (2000) attenuation curve with the {\sc MassiveBlack} predictions are shown.}
\label{fig:A1500_zevo}
\end{figure}

We now focus on the evolution of the rest-frame UV attenuation from $z\sim 10$ to $z=4-7$. In Figure~\ref{fig:A1500_zevo} we compare our predictions for the UV attenuation at $z\sim 10$ to those at $z=4-7$ applying the same methodology - using both the Meurer et al. 1999 relation and by combining predictions of the intrinsic slope with the Calzetti et al. (2000) dust curve. Assuming a constant intrinsic slopes hints at a significant increase in the UV attenuation from $z\sim 10$ to $z\sim 7$ followed by little evolution between $z=7$ and $z=4$. Utilising the intrinsic slopes predicted by {\sc MassiveBlack} (which vary with redshift) weakens this evolution.

\section{Conclusions}\label{sec:conclusions}

Here we make use of the deep {\em Hubble} and Spitzer observations available
over 5 particularly bright $z\sim10$ candidates (Oesch et al.\ 2014;
Zheng et al.\ 2012) to provide a first characterization of the mean
UV-continuum slope for a multi-object sample of galaxies at
$z\sim10$.

We find:\\

\begin{itemize}

\item Combining {\em Hubble} and {\em Spitzer} we have measured the
  mean UV continuum slope of star formation galaxy candidates at
  $z\sim 10$.  We find a mean $\beta$ of $-2.1\pm0.3\pm0.2$.  We allow
  for up to a $0.2$ error in this measurement due to systematic errors
  that may derive from the wavelength baseline used to derive $\beta$
  (Appendix A) or from small systematics in the photometry.  The
  average observed UV continuum slope of a stack of bright $z\sim 10$
  sources is only bluer than those at $z<8$ ($\beta_{\rm obs}\approx
  -1.7$) by $1\sigma$.  These measurements are more robust than those
  at $z\sim 8$ due to the wide wavelength baseline provided by the
  combination of the {\em Hubble}/WFC3 H$_{f160w}$-band and {\em
    Spitzer}/IRAC [3.6]-band.  The only previous measurement of
  $\beta$ at $z\sim10$ was by Oesch et al.\ (2014) for the most
  luminous $z\sim10$ galaxy in their selection.\\

\item These slopes are redder than the intrinsic slope predicted by galaxy formation models, suggesting the existence of some dust attenuation, even at $z\approx 10$. For the brightest candidate, GN-z10-1 (Oesch et al. 2014), we infer a rest-frame UV attenuation of $\approx 0.9$ assuming the Calzetti et al. (2000) attenuation curve ($\approx 0.5$ assuming an SMC extinction law), while for the stack of bright candidates we find $0.9\pm 0.4$ assuming the Calzetti et al. (2000) law ($0.6\pm 0.2$ assuming an SMC extinction law). \\

\end{itemize}

\subsection{Acknowledgements}

We acknowledge useful conversations with Dan Coe on this topic, who
has been attempting similar measurements on his triply imaged
$z\sim11$ candidate.  SMW acknowledges support from the Science and
Technology Facilities Council.  IL acknowledges support from the
European Research Council grant HIGHZ no. 227749 and the Netherlands
Organisation for Scientific Research Spinoza grant. The Dark Cosmology
Centre is funded by the Danish National Research Foundation.

\bsp

\appendix
\section{The effect of the wavelength baseline on the measured UV continuum slope}

The UV continua of real stellar populations, though well approximated by, does not perfectly follow a power law. This effectively leaves the power law slope $\beta$ inferred from observations sensitive to the wavelength baseline of the observations.

\begin{figure}
\centering
\includegraphics[width=20pc]{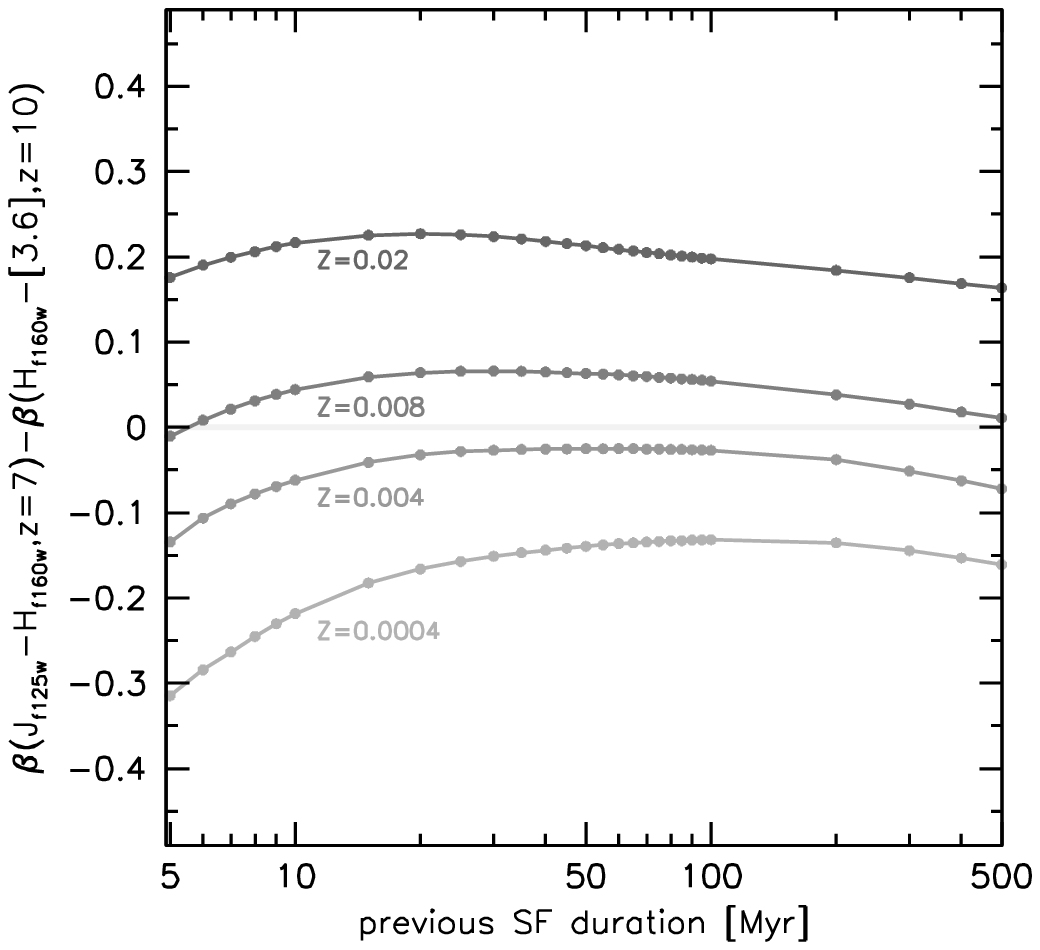}
\caption{The difference between the UV continuum slope measured at $z=7$ using the $J_{f125w}-H_{f160w}$ colour and that measured at $z=10$ using the $H_{f160w}-$[3.6] colour as a function of the previous duration of (constant) star formation for 4 different metallicities ($Z\in\{0.0004,0.004,0.008,0.02\}$.}
\label{fig:wavelength_baseline_TESTING}
\end{figure}

In Figure~\ref{fig:wavelength_baseline_TESTING} we demonstrate the difference between the UV continuum slope measured at $z=7$ using the $J_{f125w}-H_{f160w}$ colour and that measured at $z=10$ using the $H_{f160w}-$[3.6] colour for different durations of previous constant star formation and stellar metallicities. With the exception of very young durations of star formation the offset between the measured values of $\beta$ is fairly constant with star formation duration. However, the offset is strongly affected by the stellar metallicity. For example, for stellar metallicities of $Z=0.02$ there is a significant deviation between the two measures of $\beta$ with $\Delta\beta\approx 0.2$ suggesting that a much bluer value of $\beta$ would be measured at $z=10$ compared to $z=7$. However, for metallicities $Z=0.004-0.008$, which are expected to be typical of star forming galaxies at $Z=0.004-0.008$ (Wilkins et al., {\em in-prep}\footnote{Using results from the {\sc MassiveBlack-II} simulation.}) this effect becomes small ($|\Delta\beta|<0.05$), though reverses at very-low metallicity.

\end{document}